\documentstyle[11pt,newpasp,twoside,epsf]{article}
\newcommand{\beq}       {\begin{equation}}
\newcommand{\eeq}       {\end{equation}}
\newcommand{\beqa}      {\begin{eqnarray}}
\newcommand{\eeqa}      {\end{eqnarray}}

\def\avg#1              {\langle #1\rangle}     

\newcommand{\calc}      {\ifmmode {{\cal C}} \else ${{\cal C}}$\fi}
\newcommand{\calf}      {\ifmmode {{\cal F}} \else ${{\cal F}}$\fi}
\newcommand{\calg}      {\ifmmode {{\cal G}} \else ${{\cal G}}$\fi}
\newcommand{\calh}      {\ifmmode {{\cal H}} \else ${{\cal H}}$\fi}
\newcommand{\call}      {\ifmmode {{\cal L}} \else ${{\cal L}}$\fi}
\newcommand{\calm}      {\ifmmode {{\cal M}} \else ${{\cal M}}$\fi}
\newcommand{\caln}      {\ifmmode {{\cal N}} \else ${{\cal N}}$\fi}
\newcommand{\calo}      {\ifmmode {{\cal O}} \else ${{\cal O}}$\fi}
\newcommand{\calp}      {\ifmmode {{\cal P}} \else ${{\cal P}}$\fi}
\newcommand{\calq}      {\ifmmode {{\cal Q}} \else ${{\cal Q}}$\fi}
\newcommand{\calr}      {\ifmmode {{\cal R}} \else ${{\cal R}}$\fi}
\newcommand{\cals}      {\ifmmode {{\cal S}} \else ${{\cal S}}$\fi}
\newcommand{\calt}      {\ifmmode {{\cal T}} \else ${{\cal T}}$\fi}
\newcommand{\calv}      {\ifmmode {{\cal V}} \else ${{\cal V}}$\fi}
\newcommand{\calw}      {\ifmmode {{\cal W}} \else ${{\cal W}}$\fi}

\newcommand{\cm}        {\ifmmode {\rm cm}\else cm\fi}
\newcommand{\m}         {\ifmmode {\rm m}\else m\fi}
\newcommand{\km}        {\ifmmode {\rm km}\else km\fi}
\newcommand{\pc}        {\ifmmode {\rm pc}\else pc\fi}
\newcommand{\kpc}       {\ifmmode {\rm kpc}\else kpc\fi}
\newcommand{\Mpc}       {\ifmmode {\rm Mpc}\else Mpc\fi}
\newcommand{\Gpc}       {\ifmmode {\rm Gpc}\else Gpc\fi}
\newcommand{\ly}        {\ifmmode {\rm ly}\else ly\fi}
\newcommand{\Rsun}      {\ifmmode {R_\odot}\else R$_\odot$ \fi}
\newcommand{\AU}        {\ifmmode {{\rm AU}}\else AU \fi} 
\newcommand{\s}         {\ifmmode {\rm s}\else s\fi}
\newcommand{\Hz}        {\ifmmode {\rm Hz}\else Hz\fi}
\newcommand{\yr}        {\ifmmode {\rm yr}\else y\fi}
\newcommand{\cms}       {\ifmmode {\rm cm~s^{-1}}\else cm~s$^{-1}$\fi}
\newcommand{\kms}       {\ifmmode {\rm km~s^{-1}}\else km~s$^{-1}$\fi}
\newcommand{\K}         {\ifmmode {\rm K}\else K\fi}
\newcommand{\ster}      {\ifmmode {\rm ster}\else ster\fi}
\newcommand{\erg}       {\ifmmode {\rm erg}\else erg\fi}
\newcommand{\dyn}       {\ifmmode {\rm dyn}\else dyn\fi}
\newcommand{\mug}       {\ifmmode {\mu \rm{G}} \else {$\mu$G}\fi}
\newcommand{\Msun}      {\ifmmode {M}_{\mathord\odot}\else 
                          $M_{\mathord\odot}$\fi}
\newcommand{\msun}      {\ifmmode {M}_{\mathord\odot}\else 
                          $M_{\mathord\odot}$\fi}
\newcommand{\Lsun}      {\ifmmode {L}_{\mathord\odot}\else 
                          $L_{\mathord\odot}$\fi}
\newcommand{\rate}      {\ifmmode {\rm cm^3~s^{-1}}\else cm$^3$~s$^{-1}$\fi}

\newcommand{\e}         {\ifmmode ^{-1}\else $^{-1}$\fi}
\newcommand{\ee}        {\ifmmode ^{-2}\else $^{-2}$\fi}
\newcommand{\eee}       {\ifmmode ^{-3}\else $^{-3}$\fi}

\newcommand{\alphat}    {\alpha^{(2)}}
\newcommand{\mdw}       {\dot m_w}
\newcommand{\mds}       {\dot m_*}
\newcommand{\muh}       {\mu_{\rm H}}

\newcommand{\vp}        {\varpi}
\newcommand{\vpc}       {\varpi_c}
\newcommand{\vpco}      {\varpi_{c0}}
\newcommand{\vpm}       {\varpi_{\rm max}}
\newcommand{\vpmo}      {\varpi_{\rm max0}}
\newcommand{\vpo}       {\varpi_0}
\newcommand{\vzcs}      {v_{zc*}}
\newcommand{\xm}        {x_{\rm max}}
\newcommand{\xom}       {x_{0\, \rm max}}

\newcommand{\smyr}{{ M_\odot\ \rm yr^{-1}}}
\newcommand{\sm}{{ M_\odot}}

\def\ion#1#2{#1$\;${\small\rm II}\relax}
\def\lesssim{\mathrel{\hbox{\rlap{\hbox{\lower4pt\hbox{$\sim$}}}\hbox{$<$}}}}
\def\gtrsim{\mathrel{\hbox{\rlap{\hbox{\lower4pt\hbox{$\sim$}}}\hbox{$>$}}}}

\def\edcomment#1{\iffalse\marginpar{\raggedright\sl#1\/}\else\relax\fi}
\reversemarginpar

\begin{document}
\title{Outflow-Confined \ion{H}{2} Regions and the\\
Formation of Massive Stars by Accretion}
\author{Jonathan C. Tan}
\affil{Princeton University Observatory, Princeton, NJ 08544, USA}
\author{Christopher F. McKee}
\affil{Depts. of Physics \& Astronomy, UCB, Berkeley, CA 94720, USA}

\begin{abstract}
  If massive stars form by disk accretion, then bipolar outflows
  should be generated as in the case of low-mass star formation. High
  accretion rates lead to high outflow rates and make the wind density
  very large near the protostar. 
We therefore predict that massive protostars have very small,
 jet-like  \ion{H}{2} regions confined by their outflows; we
identify these \ion{H}{2} regions with ``hypercompact" ($\leq 0.01\:{\rm pc}$) 
\ion{H}{2} regions.
Their lifetime is 
approximately equal to the
  accretion timescale ($\sim10^5$~yr), much longer than the
  sound-crossing and dynamical timescales of the ionized region.
  We present an analytic description of the density distribution of
  the outflow, relate the overall mass loss rate to the accretion
  rate, and normalize to values appropriate to massive protostars.
  For a given ionizing luminosity, we calculate the extent of the \ion{H}{2}
  region and its radio spectrum. A detailed comparison is made with
  observations of radio source ``I'' in the Orion Hot Core. The spectra
  and morphologies of many other sources are broadly consistent with
  this model. We argue that confinement of ionization allows disk and
  equatorial accretion to continue up to high masses,
  potentially overcoming a major difficulty of standard accretion
  models.  
\end{abstract}

\section{Introduction}

The basic mode of massive star formation is a matter of
debate. Extensions of core collapse models (e.g. Shu, Adams, \&
Lizano 1987) from low to high masses have been criticized on the
grounds that radiative feedback is likely to disrupt accretion and
that massive stars are known to form in the crowded environments of
star clusters, where dynamical interactions between forming stars may
be important. These concerns have stimulated the development of models
that involve direct stellar collisions as well as competitive
accretion of material in proto star clusters (Bonnell, Bate, \&
Zinnecker 1998; Bonnell \& Bate 2002). However, many of the
difficulties facing standard accretion models may be resolved once the
appropriate initial conditions of massive star formation are allowed
for (McKee \& Tan 2003). In particular, the high pressures of these
regions mean that the initial cores, if close to equilibrium, will be
very dense with small volume filling factor, short collapse times, and
high accretion rates. Here we investigate the consequences of high
accretion rates on hydromagnetic outflows, and the ability of
these outflows to confine ionizing radiation from the
protostar. The radio emission from these \ion{H}{2}
regions is a useful diagnostic of the massive star formation
mechanism.  The outflow may shield the collapsing gas core from the
brunt of the protostar's radiative feedback.

\section{Protostellar Outflow Density}

We wish to determine the density distribution in the winds (we shall
use the terms winds and outflows interchangeably) from protostars and
their accretion disks.  Such winds are believed to be hydromagnetic in
origin, at least for low-mass stars.  Two classes of models have been
developed: disk winds, in which the wind flows along field lines
threading the accretion disk (e.g. K\"onigl \& Pudritz 2000), and
X-winds, in which the wind is launched from the interaction region
between the stellar magnetic field and the inner edge of the accretion
disk (e.g. Shu et al. 2000).  Here we shall develop an approximate
formulation for the density that applies to both models.  We shall not
attempt to describe the region where the wind accelerates, since that
depends on the details of the acceleration mechanism.  Instead, we
shall determine a characteristic density near the disk surface, and a
general expression for the wind density far from the disk.

\begin{figure} 
\label{fig:geom}
\plotfiddle{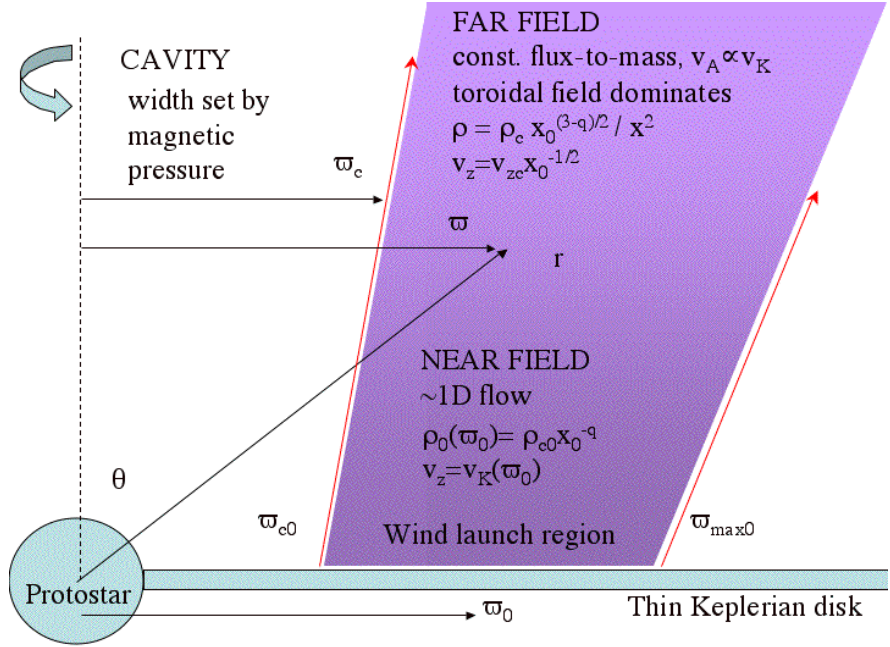}{3in}{0}{46}{46}{-170}{-15}
\caption{Schematic of hydromagnetic outflow from protostar and accretion disk. 
See text for details.}
\end{figure}

Consider a thin, Keplerian accretion disk (Fig. 1).  We
assume that the wind extends from $\vpco$ to $\vpmo$, where $\vp\equiv
r\sin\theta$ is the cylindrical radius and $\vpo$ is measured in the
plane of the disk.  Let $\rho_0(\vpo)$ be the density for the
streamline originating at $\vpo$ at the point at which the vertical
velocity $v_z$ reaches the Keplerian velocity $v_K(\vpo)$.  We assume
that $v_z$ reaches $v_K(\vpo)$ at $\vp\simeq\vpo$, and that the
density there varies as a power-law in radius,
\beq
\rho_0(\vpo)=\rho_{c0} \left(\frac{\vpo}{\vpco}\right)^{-q}
\equiv \rho_{c0}x_0^{-q}\; .
\eeq
The classic Blandford \& Payne (1982) self-similar wind has $q=3/2$;
the cylindrical self-similar wind models of Ostriker (1997) have
$1/2\leq q\leq 1$.  (Note that the Ostriker winds are prevented from
expanding laterally by an external pressure, and have magnetosonic
Mach numbers $\calm_F<1$.) 
Since the Keplerian velocity can be expressed
as $v_K=v_{Kc}x_0^{-1/2}$, the mass loss rate in the wind from both
sides of the disk is
\beq
\mdw = 4\pi\int_{\vpco}^{\vpmo} \vpo d\vpo
\rho_{0}(\vpo)v_K(\vpo)
= 4\pi\rho_{c0}v_{Kc}\vpco^2 I_{w0}\;,
\eeq   
where
$I_{w0}\equiv (\xom^{3/2 -q}-1)/(3/2 - q)$.
For an X-wind, assume that the wind emerges from a narrow
band of width $\Delta \xom$, so that $\xom=1+\Delta \xom$; then
$I_{w0}=\Delta \xom$.
For both disk winds and X-winds, the density in the wind near the disk
is then
\beq
\rho_0(\vpo)=\left(\frac{\mdw }{4\pi v_{Kc}\vpco^2I_{w0}}\right) 
x_0^{-q}\; .
\eeq
Since the flow near the disk is approximately one-dimensional, the
density at a point where the vertical velocity differs from the local
Keplerian velocity is $\rho=\rho_0[v_K(\vpo)/v_z]$.

        Next, consider the wind far from the disk. The flux-to-mass
ratio is constant along streamlines, so
$d\Phi / d\mdw= B_z / (\rho v_z)= B_{z0}/(\rho_0 v_K)$.
To simplify the discussion, we assume that the wind is self-similar,
with the Alfv\'en velocity $v_A$ proportional to the Kepler velocity
$v_K$.  In that case, $B_{z0}\propto x_0^{-(q+1)/2}$ and
\beq
\frac{B_z}{\rho v_z}=\left(\frac{B_{zc}}{\rho_c v_{zc}}\right)
        x_0^{q/2}
\label{eq:bzrhovz}
\eeq
(e.g., Ostriker 1997), where $\vp_c$ is the inner edge of the wind.
We assume that the wind has expanded so that
$\vp\gg \vpo$. In that case, the azimuthal velocity is small, and the
toroidal wrapping of the field gives 
$B_z / B_\phi = v_z /(\Omega_0 \vp)$,
where $\Omega_0$ is the angular velocity of the footpoint.
The field is approximately force free, and since the azimuthal field
dominates, we have 
\beq 
B_\phi=B_{\phi c}/x\; ,
\label{eq:bphi}
\eeq
where $x\equiv \vp/\vpc$.  We then obtain
\beq
B_z=\left(\frac{B_{\phi c} v_{zc}}{\Omega_{c0}\vpc}\right)
\frac{x_0}{x^2} 
\equiv B_{zc}\;\frac{x_0}{x^2}\; ,
\label{eq:bz}
\eeq
so that from equation (\ref{eq:bzrhovz})
\beq
\rho=\rho_c\;\frac{x_0^{(3-q)/2}}{x^2}\; .
\eeq
The result that $\rho\propto\vp^{-2}$ was first found by Shu et
al. (1995) for the particular case of X-winds, and was derived more
generally by Ostriker (1997) and by Matzner \& McKee (1999).

        Since the wind is self-similar, the velocity scales with the
footpoint velocity, $v_z=v_{zc}x_0^{-1/2}$.  Evaluating the mass loss
rate in the wind far from the disk determines the density at the
inner edge of the wind, $\rho_c$.  The density distribution far from
the disk is then
\beq
\rho  = \left(\frac{\mdw}{ 4 \pi  v_{zc}\vpc^2 I_{w\infty}}\right)
\frac{x_0^{(3-q)/2}}{x^2}\; ,
\label{eq:rhofar}
\eeq
where
$ I_{w\infty}\equiv \int_1^{\xm} dx x_0^{{1- q/2}}/x$.
We assume that $x_0=x^\beta$ for some value of $\beta$; the
coefficient in this expression is unity because the streamline
that originates at $x_0=1$ has $x=1$. Evaluating the integral,
we find
\beq
I_{w\infty}=\left(\frac{\xom^{{1-\frac 12 q}}-1}{1-\frac 12 q}\right)
        \frac{\ln\xm}{\ln\xom}.
\eeq
Since X-winds emanate from a narrow range of radii, we set $x_0\simeq 1$ for
them; X-winds therefore have $I_{w\infty}=\ln \xm$.

        Combining these results, we express the density in the wind as
\beq
\rho=\frac{\mdw f(x_0,x)}{4\pi v_{zc}\vpc^2 I_w},
\label{eq:rho}
\eeq
where
\beq
f(x_0,x)=\cases{
        \displaystyle x_0^{-q} & $x\simeq x_0$ , \cr
        \displaystyle \frac{x_0^{(3-q)/2}}{x^2} & $x\gg x_0$ .
        \cr}
\eeq
At the disk surface, $v_{zc}=v_{Kc}$ and  $I_w=I_{w0}$,
whereas far from the disk $v_{zc}>v_{Kc}$ and $I_w=I_{w\infty}$.
Note that for $q=1$, we have $f=x_0/x^2$ in both cases, so
that in this case
\beq
I_w=2(\xom^{1/2}-1)\left(\frac{\ln \xm}{\ln\xom}\right)~~~~~(q=1),
\label{eq:iw}
\eeq
both at the disk and far from the disk.
For X-winds, $I_{w\infty}=\ln\xm$ reduces
to $I_{w0}=\Delta \xom$ at the disk surface, so we can take
$I_w\simeq \ln\xm$ everywhere for X-winds.
    
    We evaluate equation (\ref{eq:rho}) numerically in terms of
the density of hydrogen nuclei, $n_w\equiv\rho/\muh$, where
$\muh=2.34\times 10^{-24}$ g is the mean mass per hydrogen for an
assumed helium abundance 10\% of hydrogen.  We also write the 
mass loss rate in the wind as $\mdw\equiv f_w\mds$, where $\mds$ is
the accretion rate onto the star. For example, Matzner \& McKee (1999)
estimate $f_w\simeq \frac 15$.  We then find
\beq
n_w\equiv\frac{\rho}{\muh}=9.52\times 10^{10}\left[\frac{f_w f(x_0,x)}{
        v_{zc,7}I_w}\right]\left(\frac{\mds}{10^{-4}~M_\odot~
        \yr\e}\right)\left(\frac{1~{\rm
        AU}}{\vpc}\right)^2~~~~~\cm^{-3} \; ,
\eeq
where $v_{zc,7}\equiv v_{zc}/(10^7\ \cms)$.

\subsection{Axial Wind Cavity}

A question of critical importance for \ion{H}{2} regions associated
with massive protostars is the shape of the cavity that forms along
the axis, since ionizing photons can propagate freely through this
cavity.  There are several effects that govern its shape:
For disk winds, the flow is launched at an angle greater than
$30\deg$ from the rotation axis, and the hoop stresses of the toroidal
field must overcome the inertia of the outflow in order to begin to
close the cavity. For X-winds, poloidal field lines from the protostar
determine the shape of the cavity (Shu et al. 1995), and the same
appears to be true for disk winds (E. Ostriker 2002, private comm.).
In either case, once the protostar becomes massive enough to generate
a strong main sequence wind, the pressure due to this wind will tend
to open the cavity further.

For simplicity, we shall focus on X-winds, for which $x_0=1$ as
discussed above.  The magnetic flux in the cavity is $\Phi_c=\pi
B_c\vpc^2$, whereas that in the wind far from the disk can be
evaluated with the aid of equation (\ref{eq:bz}),
\beq
\Phi_w=\frac{2\pi \vpc v_{zc} B_{\phi c}}{\Omega_{c0}}\;
\ln\frac{\vpm}{\vpc}\; .
\eeq
Since the maximum
extent of the wind $\vpm$ enters only logarithmically, we approximate
it as $\vpm\simeq r_c $.
Let $\vzcs\equiv v_{zc}/v_{Kc}=v_{zc}/\Omega_{c0}\vpco$ 
be the normalized velocity;
in the numerical example worked out by Shu et al. (1995), $\vzcs=2.1$.
Magnetic pressure balance requires 
that the field in the cavity, $B_c$, equal the field at the inner edge
of the wind, $B_{\phi c}$ (which is much greater than $B_z$ there).  
Furthermore, in the X-wind model, the
cavity flux and the wind flux are comparable, $\Phi_c\simeq \Phi_w$.
These relations give an equation for the radius of the 
cavity,
\beq
\vpc= 2\vzcs \vpco\; \ln\frac{r_c}{\vpc}\; ,
\label{eq:vpc}
\eeq
which is valid far from the disk.  
We develop an approximate solution of equation (\ref{eq:vpc})
following the approach of Matzner \& McKee (1999), and then add
a term in the logarithm to make it valid at the disk surface:
\beq
\vpc\simeq 1.4\vzcs\vpco \ln\left[\frac{r_c}
{1.4\vzcs\vpco}+e^{1/(1.4\vzcs)}-\frac{1}{1.4\vzcs}\right]\; .
\label{eq:vpcapprox}
\eeq
This form remains valid for regions near the disk even when $\vzcs\neq 1$.

        We now evaluate the factor 
$\ln \xm$ in $I_{w\infty}$, which
enters the expression for the density far from the disk:
\beq
\ln\xm=\ln\frac{\vpm}{\vpc}\simeq\ln\frac{r_c}{\vpc}\simeq
0.7\ln\left(\frac{r_c}
{1.4\vzcs\vpco}\right)\; ,
\eeq
where we have assumed $r_c\gg \vpco$.  We generalize to arbitrary
values of $r_c$ by adding a term in the logarithm,
\beq
\ln\xm\simeq 0.7\ln\left(\frac{r_c}{1.4\vzcs\vpco}+\xom^{1/0.7}-\frac{
             1}{1.4\vzcs}\right)\; .
\label{eq:lnxmapprox}
\eeq
Our result for the density agrees with that obtained
by Shu et al. (1995) to within about 20\% for $10\la r_c/\vpco\la 10^5$.

\section{Outflow-Confined \ion{H}{2} Regions and Application in Orion}

\begin{figure} 
\plottwo{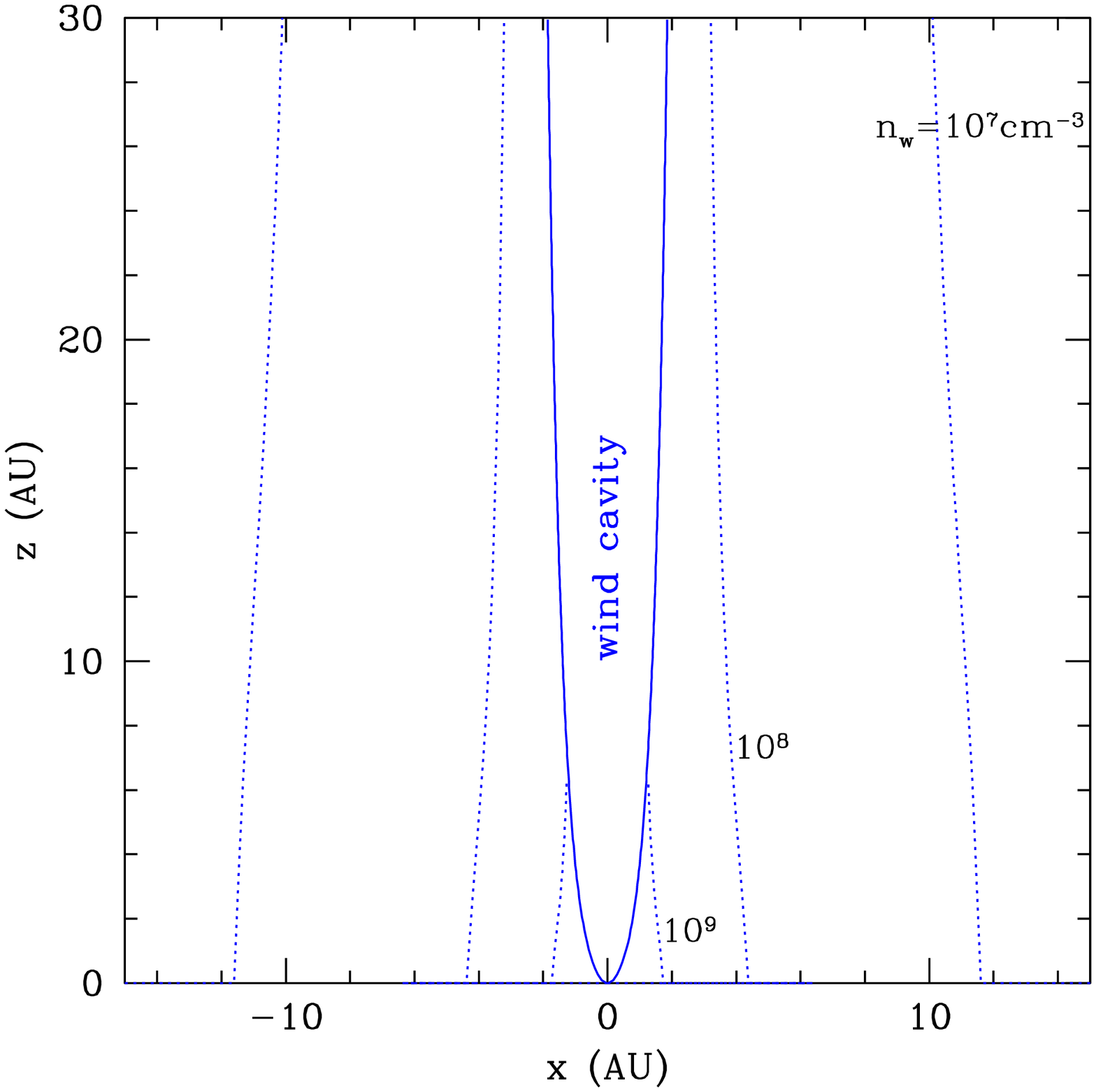}{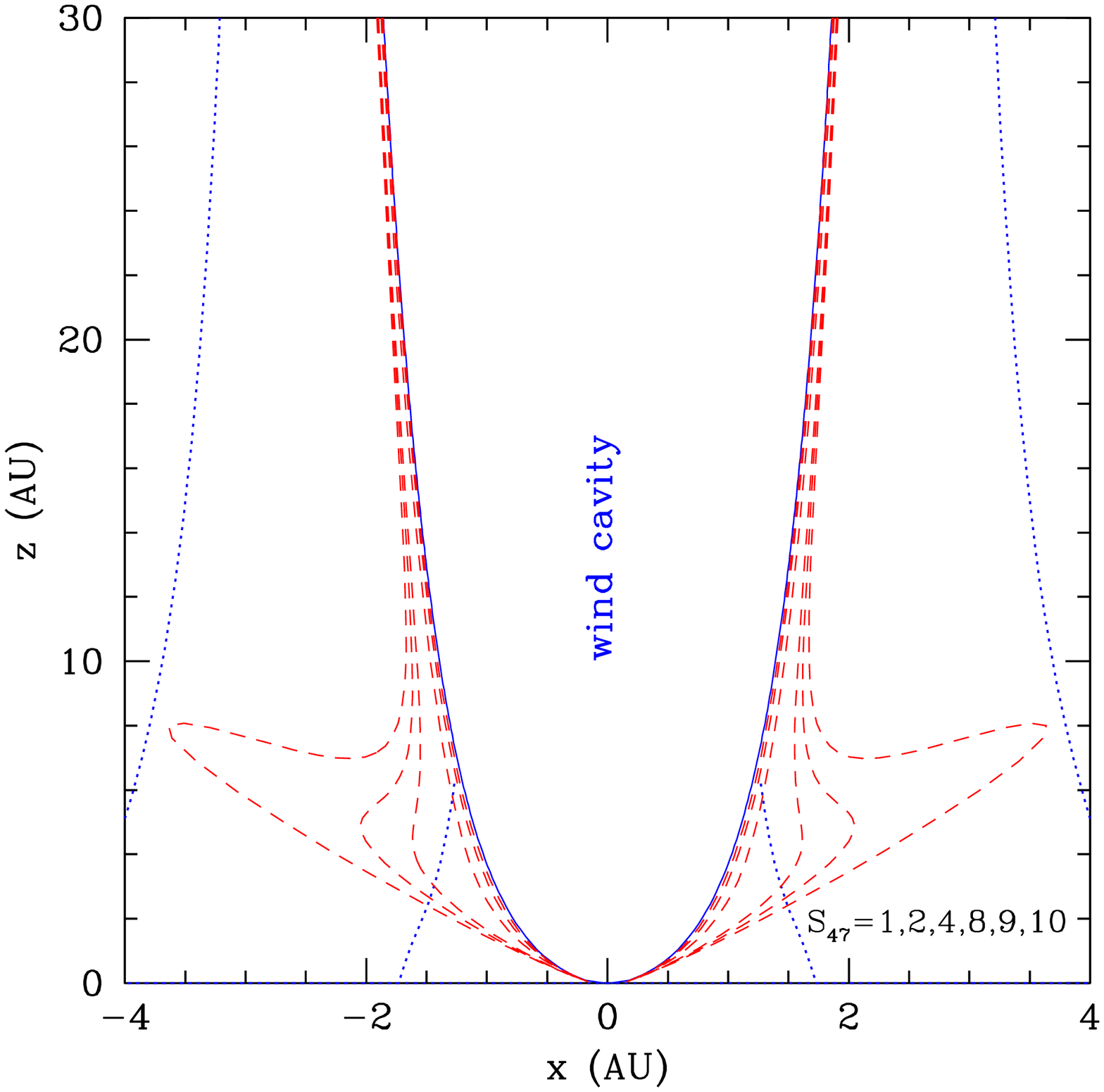}
\caption{Left: Outflow isodensity contours from a $20\sm$ protostar, with $r_*=16R_\odot$, $\vpco=2r_*$, $\mds=10^{-4}\smyr$, $\dot{m}_w=\dot{m}_*/3$, $v_{zc*}=2.1$, and $q=1$. Right: \ion{H}{2} region growth with ionizing luminosity, $S$.}
\end{figure}

        Consider a beam of ionizing radiation from the protostar that
intercepts the cavity boundary at a distance $r_c(\theta)$.  
Let $S$ be the ionizing photon luminosity, so that the rate
at which ionizing photons cross an element of 
area $dA$ 
is $SdA/4\pi r_c^2$.
This flux of ionizing photons is able to balance the recombinations
of the ions in the wind out a radius $r_i$
given by
\beq
\frac{S dA}{4\pi r_c^2}=\frac{dA}{r_c^2}\int_{r_c}^{r_i}
\alphat n_w^2r^2 dr\; 
\label{eq:sda}
\eeq
where $\alphat$ is the radiative recombination rate to the excited
states of hydrogen and the gas is assumed to be fully ionized inside
$r_i$.  This equation ignores the influx of neutrals from the
wind and the absorption of ionizing photons by dust, which both make
the \ion{H}{2} region smaller. However, the effect of neutral influx
is negligible for the specific numerical models we consider for the
Orion source (below).  If the outflow originates from close to the
protostar then most dust grains are likely to have
been destroyed.  We present an analytic description of the \ion{H}{2}
region elsewhere. For now we solve equation (\ref{eq:sda})
numerically (Fig.~2).

For small ionizing luminosities only a thin skin of material around
the cavity is ionized. As the luminosity increases the \ion{H}{2}
region expands out to large angles from the rotation axis and greater
distances from the protostar. At a critical ionizing luminosity the
\ion{H}{2} region breaks out at angles $\sim 30\deg$ from the rotation
axis. Soon the entire solid angle interior to this becomes ionized,
while in the equatorial directions the gas can remain neutral for a
much longer time.

We now apply this model to the massive protostar powering the Orion
hot core.  At a distance of 450~pc, this is the closest example of a
massive star in formation. The core is self-luminous ($L_{\rm bol}\sim
1-5\times 10^4\:{\rm L_\odot}$, Gezari et al. 1998; Kaufman et al.
1998). A weak radio continuum source (``I'') (e.g.  Menten \& Reid
1995), located within a few arcseconds of the core center, as traced
by dust and gas emission (Wright, Plambeck \& Wilner 1996), almost
certainly pinpoints the location of the massive protostar. SiO
emission forms a ``bow-tie'' feature centered on ``I'', that may be an
inclined or flared disk (Wright et al. 1995).  Perpendicular to this,
a large scale, wide-angle bipolar outflow extends to the NW and SE of
the core (Chernin \& Wright 1996; Greenhill et al. 1998). Chernin \&
Wright modeled the flow axis as being inclined at $65^\circ$ to our
line of sight (Fig 3a). At 22~GHz, source ``I'' appears elongated
(0\arcsec.145 by $<0\arcsec.085$, Menten 2002, private comm.) parallel
to the large scale outflow axis.

We model the thermal bremsstrahlung emission from source ``I''
assuming we are viewing the outflow at $\theta_{\rm view}=65\deg$ and
that the temperature of the ionized gas is $10^4\:{\rm K}$. As we
assume negligible density in the wind cavity, the \ion{H}{2} region is
formally infinite in length. However, the emission measure rapidly
decreases along the jet (as $r^{-3}$). To compare to observations at
8.4~GHz (Menten \& Reid 1995), 14.9~GHz (Felli et al. 1993), 22.3~GHz
(Menten 2002, private comm.), 43.1~GHz (Menten \& Reid 1995), and
86~GHz (Plambeck et al. 1995), we include emission from scales about
one third larger than the quoted beam sizes, i.e.  0.3\arcsec,
0.2\arcsec, 0.2\arcsec, 0.3\arcsec, and 0.5\arcsec, respectively. We
extend to higher frequencies with the 0.5\arcsec size. Upper limits at
98~GHz and 218~GHz are from Murata et al. (1992) and Blake et al.
(1996), respectively. A more detailed comparison of predicted surface
brightness profiles with data will be presented elsewhere.
The radiative transfer calculation is accomplished by dividing the
\ion{H}{2} region, which is always axisymmetric in our modeling, into
many volume elements and then calculating optical depths along ray
paths to the observer. We increase the resolution to achieve convergence.

Our fiducial protostellar model is based on constraints derived from
the total luminosity (McKee \& Tan 2003). We set $m_*=20\sm$ and
$\mds=10^{-4}\smyr$. We assume $\dot{m}_w=0.33\mds$ and $v_w=2.1v_K$,
which are reasonable choices for certain classes of disk winds and
X-winds. Given the uncertainties in the ionizing luminosity of massive
protostars, we consider a range of values (Fig~3b). A protostar with
$S=2\times 10^{47}\:{\rm s^{-1}}$ provides a
good match to the observations, and this luminosity is consistent with
the predictions of models of massive protostellar evolution (Tan 2002).

While not unique, the model we have presented for the protostar in the
Orion hot core has a bolometric luminosity consistent with infrared
observations, an accretion rate that is based on theoretical
expectations for massive gas cores collapsing in high pressure
environments, a protostellar size that is calculated given this
accretion rate, and an ionizing luminosity consistent with theoretical
models and able to create an \ion{H}{2} region with the
observed radio spectrum. The geometry of the system was chosen so that
this inner protostellar outflow naturally connects with the larger
scale flow from OMC-1.

However, the model is nonetheless highly idealized: we have assumed that the
density of material in the wind cavity is negligible, ignored the
possible influence of a conventional stellar wind on the cavity
geometry, treated the propagation of ionizing photons along sectors as
being independent, ignored dust (although much of this may have been
destroyed in the protostellar radiation field), and made simple
parameterizations for the connection between accretion and outflow
generation. Some of these issues will be addressed in a future paper.

\begin{figure} 
  \plottwo{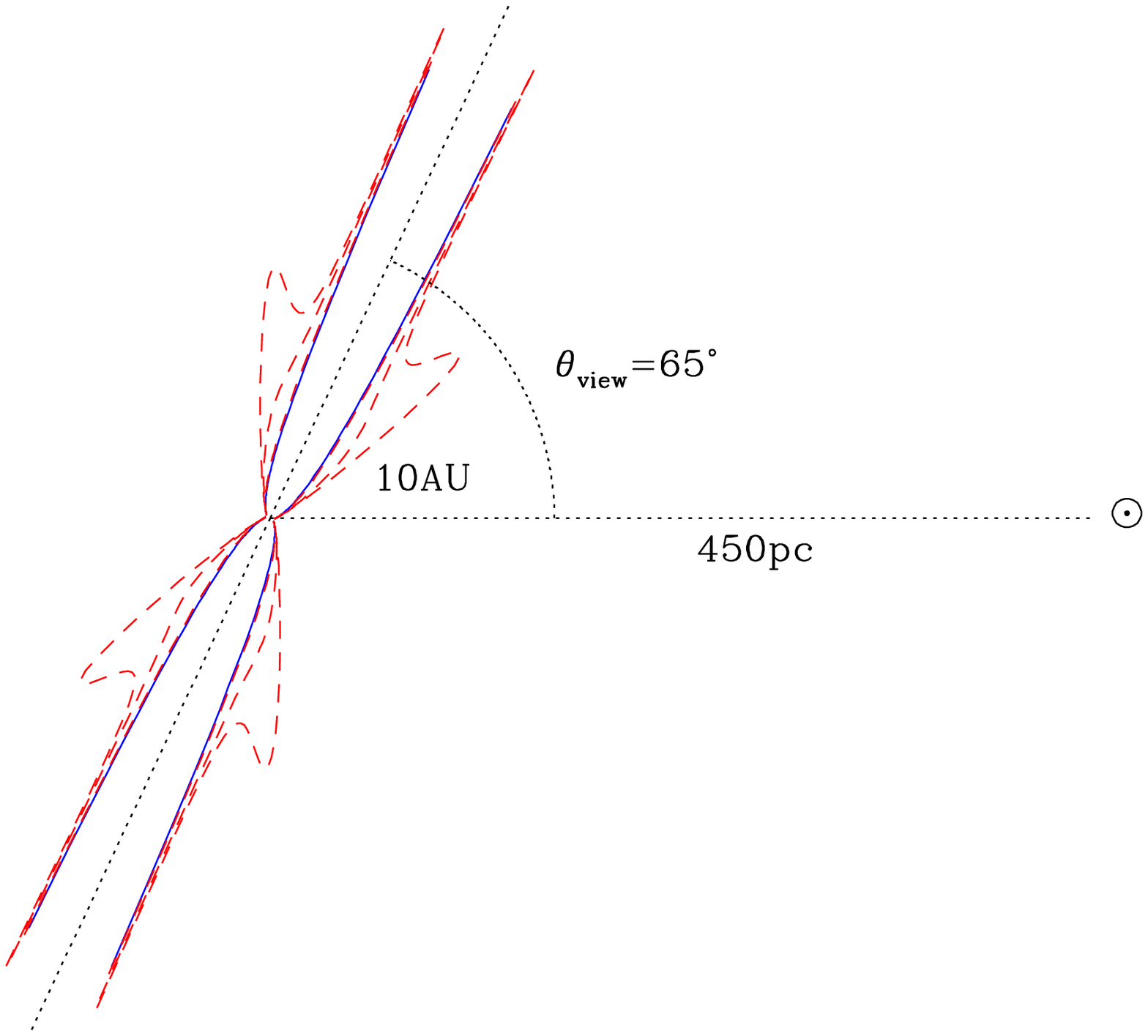}{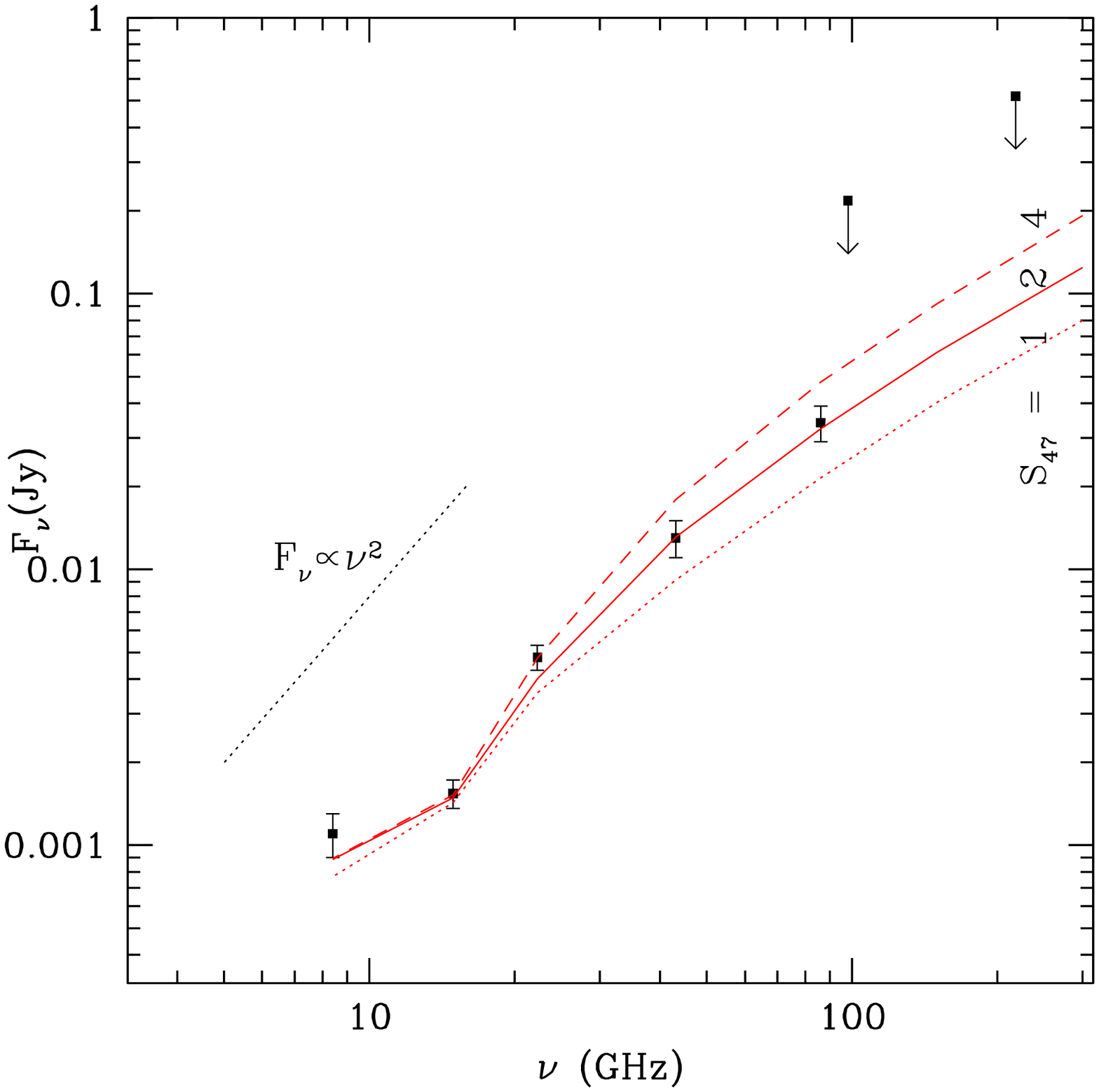}
\caption{(a) Geometry of outflow and \ion{H}{2} region. (b) Radio spectra,
  modulated by variations in observational beam size (see text), of thermal
  bremsstrahlung emission from \ion{H}{2} regions with $S_{47}=1,2,4$.}
\end{figure}

\section{Implications for Massive Star Formation}

The class of {\it outflow-confined \ion{H}{2}
  regions} is expected to be relevant to many, if not all,
massive protostars. During the earlier stages of evolution the
\ion{H}{2} region is confined close to the jet axis, resulting in a
relatively weak radio source, such as source ``I'' in Orion. At
somewhat later stages the \ion{H}{2} region is able to extend to
greater distances, although the peak of the emission measure is
expected to remain concentrated close to the protostar. This
evolutionary sequence is relevant to a host of observed sources, such as
Cepheus A~HW2 (Torrelles et al. 1996), 
IRAS~16547-4247 (Garay et al. 2003), 
IRAS~18162-2048 (G\'omez et al. 2003), 
IRAS~20126+4104 (Zhang, Hunter, \& Sridharan 1998), 
G192.16-3.82 (Shepherd, Claussen, \& Kurtz 2001),
AFGL~2591 (Trinidad et al. 2003), and
AFGL~5142 (Zhang et al. 2002).
In many of these systems jet-like morphology is clearly observed.

The small observed sizes of the \ion{H}{2} regions may be due either
to the limited sensitivity of particular observations, or to a real
physical confinement by the outflow. In the latter case the sound
crossing time (e.g. $t_s = r_{\rm HII}/c_s = 1000 [r_{\rm
  HII}/0.01\:{\rm pc}][c_s/10\kms]^{-1}\:{\rm yr}$) and flow crossing times
(typically $\sim 100$ times shorter than $t_s$) are both unrelated to
and much shorter than the true lifetime of the system, which is set by
the accretion timescale of the protostar ($\sim 10^5\:{\rm yr}$, McKee
\& Tan 2002) and the timescale for evolution of the ionizing
luminosity. This bears upon estimates of the expected number of
observed sources, in a similar manner to that of the classic
``lifetime'' problem of ultracompact \ion{H}{2} regions (Wood \&
Churchwell 1989).

One prediction of the outflow-confined model for compact \ion{H}{2}
regions is the presence of very broad recombination lines. The outflow
velocities from massive protostars are expected to be at least of
order the escape velocity, $v_{\rm esc} = 620 (m_*/10\sm)^{1/2}
(r_*/10R_\odot)^{-1/2}\kms$. Our fiducial model for source ``I'' in
Orion predicts that the fastest velocities are about $1000\kms$.
Atomic line profiles from HH objects much further out in the outflow
from OMC-1 show line-of-sight widths of several hundred $\kms$ (Taylor
et al. 1986), which would be consistent with the presence of a
$\sim1000\:\kms$ flow from the protostar. Of course as this flow
interacts with nearby dense molecular gas, it will create outflows
with a wide range of lower velocities, as are observed (Chernin \&
Wright 1996; Stolovy et al. 1998).

Outflow-confined \ion{H}{2} regions have broader implications for the
massive star formation paradigm. First of all, their properties are a
quantitative diagnostic of the accretion scenario for massive star
formation: radio spectra allow estimation of the outflow density near
the protostar. Theoretical models of outflows from accretion disks and
protostars then allow a direct connection between these observations
and the accretion physics. If the morphologies of outflows very close
to the protostar align with larger scale features, then this would
suggest that massive stars form from accretion disks that are
relatively stable in terms of their orientation. Stellar collisions
would be expected to disrupt such configurations.  We would argue that
the current observational evidence, particularly in the Orion hot
core, supports formation models based on accretion rather than stellar
collisions.

High accretion rates to massive protostars naturally result in high
outflow rates, which create dense and collimated gas structures
close to the star. These in turn should help to shield the bulk of the
accreting material from radiative protostellar feedback. In
particular, ionizing photons are degraded in the \ion{H}{2} region,
never reaching the disk in the equatorial plane. Models of disk
photoevaporation (Hollenbach et al. 1994) are therefore inhibited
during the main accretion phase. By the time the protostar has reached
a stage where the ionizing luminosity is important ($m_*\gtrsim
10\sm$) we expect the outflow to have significantly altered the
density distribution of the initial gas core in its polar regions (Matzner
\& McKee 2000). Radiant energy from the protostar will thus tend to
escape along these directions, rather that in the equatorial plane.
This is similar to the so-called ``flashlight effect'' discussed by Yorke
\& Bodenheimer (1999), although they considered that arising only from
the presence of a disk. Thus radiation pressure feedback on infall may be
substantially weakened, to an extent greater than that resulting from
rotation alone (Nakano 1989; Jijina \& Adams 1996). The evolution of
the size of the outflow-enhanced flashlight effect may
play an important role in determining the maximum stellar mass.

\acknowledgments We thank Karl Menten and Dick Plambeck for sharing unpublished results
on Orion and Eve Ostriker for helpful discussions.  
The research of JCT is supported by a Spitzer-Cotsen
fellowship from Princeton University and by NASA grant NAG5-10811. The
research of CFM is supported by NSF grant AST-0098365 and by a NASA
grant funding the Center for Star Formation Studies.

\end{document}